\documentclass{aa}  
\usepackage{graphicx}
\usepackage{caption}
\usepackage{subcaption}
\usepackage[varg]{txfonts}
\usepackage[table]{xcolor}
\usepackage{siunitx}
\usepackage{booktabs, cellspace, multirow}
\usepackage{orcidlink}
\usepackage{natbib,twoopt}
\bibpunct{(}{)}{;}{a}{}{,} % to follow the A&A style
\usepackage{comment}
\usepackage[]{hyperref}

\definecolor{cobalt}{rgb}{0.06, 0.2, 0.65}
\hypersetup{
  colorlinks,
  citecolor=cobalt,
  linkcolor=[rgb]{0.8, 0.2, 1.0},
  urlcolor=cobalt,
}
\makeatletter
  \newcommandtwoopt{\citeads}[3][][]{\href{http://adsabs.harvard.edu/abs/#3}%
    {\def\hyper@linkstart##1##2{}%
     \let\hyper@linkend\@empty\citealp[#1][#2]{#3}}}
  \newcommandtwoopt{\citepads}[3][][]{\href{http://adsabs.harvard.edu/abs/#3}%
    {\def\hyper@linkstart##1##2{}%
     \let\hyper@linkend\@empty\citep[#1][#2]{#3}}}
  \newcommandtwoopt{\citetads}[3][][]{\href{http://adsabs.harvard.edu/abs/#3}%
    {\def\hyper@linkstart##1##2{}%
     \let\hyper@linkend\@empty\citet[#1][#2]{#3}}}
  \newcommandtwoopt{\citeyearads}[3][][]%
    {\href{http://adsabs.harvard.edu/abs/#3}
    {\def\hyper@linkstart##1##2{}%
     \let\hyper@linkend\@empty\citeyear[#1][#2]{#3}}}
\makeatother

\def\gaia{\textit{Gaia}\xspace}
\def\g{$G$\xspace}

\begin{document} 
% (\gaia AGN Survey of Periodic variability)

\title{A search for periodic AGN variability in \gaia Data Release 3}
\titlerunning{A search for periodic AGN variability in \gaia Data Release 3}
\authorrunning{P. Huijse et al.}

\author{
P.~Huijse\orcidlink{0000-0003-3541-1697}\inst{1, 2}
\and
J.~Davelaar\orcidlink{0000-0002-2685-2434}\inst{3,4}
\and 
J.~De Ridder\orcidlink{0000-0001-6726-2863}\inst{1}
\and
N.~Jannsen\orcidlink{0000-0003-4670-9616}\inst{1,5}
\and
C.~Aerts\orcidlink{0000-0003-1822-7126} \inst{1,6,7}
}

\institute{
Institute of Astronomy, KU Leuven, Celestijnenlaan 200D, B-3001 Leuven, Belgium\\
\email{pablo.huijse@kuleuven.be}
\and
Millennium Institute of Astrophysics, Nuncio Monsenor Sotero Sanz 100, Of. 104, Providencia, Santiago, Chile 
\and
Department of Astrophysical Sciences, Peyton Hall, Princeton University, Princeton, NJ 08544, USA
\and
NASA Hubble Fellowship Program, Einstein Fellow
\and
Isaac Newton Group of Telescopes (Roche de Los Muchachos), Apartado 321, E-38700 Santa Cruz de La Palma, Canaries, Spain
\and
Department of Astrophysics, IMAPP, Radboud University Nĳmegen, PO Box 9010, 6500 GL Nĳmegen, The Netherlands
\and
Max Planck Institute for Astronomy, Koenigstuhl 17, 69117 Heidelberg, Germany
}

\date{Received ...; accepted ...}

\abstract
{Supermassive black hole binaries (SMBHB) are expected to produce periodic modulations in active galactic nuclei (AGN) light curves, but distinguishing such signals from stochastic red-noise variability remains a major challenge.} % context
{We present the first systematic search for statistically significant AGN periodicities 
using the optical photometry from the \gaia space mission
Data Release 3 (DR3), with the goal of identifying SMBHB candidates and establishing a 
methodological data analysis framework that can be scaled to the forthcoming Data Release~4 (DR4).}% aims
{We analyse \gaia \g band light curves of 377,128 sources from the \gaia celestial reference frame (CRF3). Stochastic variability is modelled as a damped random walk Gaussian process, and empirical false alarm probabilities are derived by comparing observed Lomb--Scargle periodogram peaks against 100,000 synthetic red-noise realisations. Candidates from this first stage are then re-evaluated using full Markov chain Monte Carlo inference under both exponential and powered-exponential kernels.} % methods
{We find 13 sources surviving our statistical criterion ($p < \alpha = 10^{-5}$) after both stages of filtering, which is consistent with the expected false-positive rate. All candidates cover fewer than 2.5 cycles of the candidate period and are systematically concentrated in a region of the parameter space indicative of model misspecification.}
% results
{No reliable periodic SMBHB candidates are retained. The ${\sim}950$-day baseline of \gaia DR3 confines all detections to the few-cycle regime where red noise most convincingly mimics periodicity, a limitation that photometric precision alone cannot overcome. The longer baseline of gaia DR4 will be essential to push beyond this regime. We offer our data analysis software pipeline in open access to the community.}
% conclusions 

\keywords{
(Galaxies:) quasars: supermassive black holes -- 
Methods: numerical -- 
Methods: statistical --
Methods: data analysis --
Techniques: photometric
}

\maketitle
\nolinenumbers

%%%%%%%%%%%%%%%%%%%%%%%%%%%%%%%%%%%%%%%%%%%%%%%%%
\section{Introduction}
%%%%%%%%%%%%%%%%%%%%%%%%%%%%%%%%%%%%%%%%%%%%%%%%%

Active galactic nuclei (AGN) are the luminous cores of galaxies powered by accretion onto a supermassive black hole. This accretion process is inherently dynamic. AGN are observed to vary in flux across the electromagnetic spectrum on timescales ranging from hours to years, with optical variability generally well described by a stochastic red-noise process \citep{kelly2009variations, kozlowski2009quantifying, macleod2010sdssdrw}. Superimposed on this stochastic behaviour, periodic or quasi-periodic signals have been reported in a number of sources. One particularly compelling origin for strictly periodic variability is the presence of supermassive black hole binaries (SMBHBs), which form through galaxy mergers \citep{begelman1980smbhb} and are expected to modulate AGN emission at the orbital period through accretion rate variations, relativistic Doppler boosting, or gravitational self-lensing \citep{shi2012accretion,dorazio2013accretion,dorazio2015dopplerboost,Hu2020,davelaar2022}.

Over the past decade, several systematic searches for periodicities in large samples of AGN light curves have been carried out using time-domain surveys. \citet{graham2015systematic} analysed 243\,500 spectroscopically confirmed quasar light curves from the Catalina Real-time Transient Survey (CRTS), identifying 111 candidates. Among the most notable results from CRTS, \citet{graham2015possible} reported a strong sinusoidal signal with a mean observed period of $1\,884 \pm 88$ days in the quasar PG~1302-102, as evidence for a close SMBHB. Subsequent work, however, casted doubt on this interpretation. \citet{vaughan2016false} demonstrated that such a period, spanning approximately twice the duration of the light curve, can arise naturally from stochastic variability with a bending power-law spectral density. \citet{liu2018asassnsmbhb} showed that the evidence for periodicity weakened when the baseline was extended by combining CRTS with ASAS-SN data, contrary to what is expected for a periodic process. More recently, \citet{zhu2020unambiguous} used Bayesian model comparison to show that, while the data support a sinusoidal component, a quasi-periodic model is statistically preferred.

Using data from the Palomar Transient Factory (PTF), \citet{charisi2016population} searched 35\,383 spectroscopically confirmed quasars and identified 50 candidates with significant periodicity, a number that reduced to 33 when PTF light curves were supplemented with CRTS observations. \citet{liu2019smbhbpanstarrs} carried out a similar effort with Pan-STARRS1 (PS1), analysing approximately 9\,000 colour-selected quasars and finding 26 significant periodicities, which upon reanalysis with an extended baseline reduced to a single candidate. \citet{Chen2020} searched for periodic variability among 625 spectroscopically confirmed quasars using 20-yr multi-colour time series constructed by combining Dark Energy Survey and Sloan Digital Sky Survey data, identifying five candidates, one of which was further investigated by \citet{Liao2021}. \citet{Chen+2024} examined Zwicky Transient Facility (ZTF) light curves of 143\,700 confirmed quasars, identifying 86 candidates that reduced to three upon extending the ZTF baseline with CRTS, PTF, PS1, and ATLAS observations. Most recently, \citet{Luo2025} analysed 48\,932 AGN using mid-infrared data from the Wide-field Infrared Survey Explorer (WISE), finding 28 candidates, including one exhibiting significant periodicity in both infrared and optical bands.

These searches have collectively demonstrated both the promise and the difficulty of identifying genuine periodicities in AGN 
light curves. \gaia DR3 \citep{vallenari2023gaia} offers a 
compelling dataset for such searches, with well-calibrated, space-based photometry for over six million QSO candidates 
\citep{bailer2023gaia, carnenero2023gaia}, approximately one million of which have publicly available epoch photometry. However, as demonstrated by \citet{El-Badry_2026}, preliminary periodic candidates identified in \gaia DR3 without adequate red-noise modelling are overwhelmingly false positives, underscoring the need for robust significance estimation. 

In this work, we present a systematic search for periodicities in the \gaia DR3 AGN sample using empirical false alarm probabilities calibrated 
against the stochastic variability of each source. The forthcoming \gaia DR4, expected by the end of 2026, will double the DR3 observation baseline and is anticipated to increase the number of published light curves by orders of magnitude. As the main aim of our work, the methodology established here is designed to scale directly to this forthcoming \gaia DR4. While designed for \gaia data analysis, it is readily applicable to other ongoing and future photometric surveys of SMBH(B)s.

%The paper is organised as follows. Section~\ref{sec:data} details the compilation of our initial AGN sample from \gaia DR3. Section~\ref{sec:methods} outlines the methodology used to identify sources that exhibit significant periodic variability. Finally, Section~\ref{sec:results} presents the results of the analysis.

%%%%%%%%%%%%%%%%%%%%%%%%%%%%%%%%%%%%%%%%%%%%%%%%%
\section{Sample selection and \gaia DR3 data}
\label{sec:data}
%%%%%%%%%%%%%%%%%%%%%%%%%%%%%%%%%%%%%%%%%%%%%%%%%

We begin our search for SMBHB candidates from the 835\,996 sources from the DR3 \texttt{qso\_candidates} table that have publicly available epoch photometry and that appear in the \gaia celestial reference frame \citep[CRF3;][]{klioner2022gaia}. This ensures that each source is present in at least one of the 17 quasar and AGN catalogues crossmatched to construct the CRF3. We apply a conservative redshift filter based on the estimates from the Quasar Classifier (QSOC) module, excluding 5\,585 sources with $z<0.1$ and 12\,532 sources lacking redshift information.

As shown by \citet{holl2023gaia}, \gaia's scanning law can introduce spurious periodicities in the photometric time series. Using two diagnostics from that work, the Spearman correlation and the significance of the scan-angle signal, we identify and remove 180 sources whose variability is significantly affected by scan-angle-dependent systematics. Finally, to ensure sufficient data for statistical modelling, we restrict the sample to sources with light curves spanning at least 500~days and containing at least 40 observations. These selection criteria yield a final sample of 377,128 sources\footnote{See Appendix~\ref{sec:supporting_material} for an ADQL query to retrieve the source identifiers and relevant metadata.}.

The light curves for this sample are retrieved using the \gaia Datalink service through the \texttt{astroquery} Python package \citep{ginsburg2019astroquery}. Observation times and magnitudes are extracted from the \texttt{g\_transit\_time} and \texttt{g\_transit\_mag} arrays in the DR3 \texttt{epoch\_photometry} table. Magnitude uncertainties are not provided directly but are derived from the \texttt{g\_transit\_flux\_over\_error} array by propagating the flux-to-magnitude conversion. Observations flagged as spurious are removed using the \texttt{variability\_flag\_g\_reject} array\footnote{This flag results from the operators described in Section~10.2.3 of the \gaia DR3 documentation \citep{rimoldini2022gaiadr3manual}.}. Even after this filtering, strong outliers may persist. We therefore apply additional cleaning: observations with magnitudes falling outside a Tukey fence\footnote{A Tukey fence 
defines outliers as values lying outside the interval $[Q_1 - k\,\mathrm{IQR},\; Q_3 + k\,\mathrm{IQR}]$, where $Q_1$ and $Q_3$ are the first and third quartiles, $\mathrm{IQR} = Q_3 - Q_1$, and $k$ is a chosen multiplier.} with $k=1.5$ are excluded. Observations with magnitude errors exceeding either an absolute threshold of 0.5~mag or an upper Tukey fence with $k=3$ are also removed.

%%%%%%%%%%%%%%%%%%%%%%%%%%%%%%%%%%%%%%%%%%%%%%%
\section{Methodological data analysis pipeline}
%%%%%%%%%%%%%%%%%%%%%%%%%%%%%%%%%%%%%%%%%%%%%%%
\label{sec:methods}

Irregular sampling and heteroscedasticity in time-series data complicate the application of standard Fourier-based methods commonly used to search for periodic phenomena in astronomical light curves.
Among the alternatives developed for this setting, the Lomb–Scargle (LS) periodogram \citep{lomb1976least, scargle1982studies, vanderplas2018understandingls} is perhaps the most widely adopted. The significance of a periodogram frequency peak is commonly assessed through false alarm probabilities (FAPs), which typically assume white Gaussian noise as the null hypothesis \citep{baluev2008fap}.
AGN, however, exhibit stochastic variability across a broad range of timescales that is better characterised as red noise, where power increases toward lower frequencies, rendering conventional FAPs unreliable \citep{vaughan2016false}. 

Several studies have found the damped random walk (DRW) model\footnote{Also referred to as the Ornstein–Uhlenbeck process.} to provide a reasonable statistical description of AGN variability \citep{kelly2009variations, kozlowski2009quantifying, macleod2010sdssdrw}, although its limitations have been documented \citep{zu2013isquasardpw, kozowski2017limitationsdrw}. Despite these caveats, the DRW remains standard practice in AGN periodicity searches \citep{charisi2016population, liu2019smbhbpanstarrs, witt2022periodicquasars, davis2024reliable, Luo2025}.

\begin{figure*}[t]
    \centering
    \includegraphics[width=\textwidth]{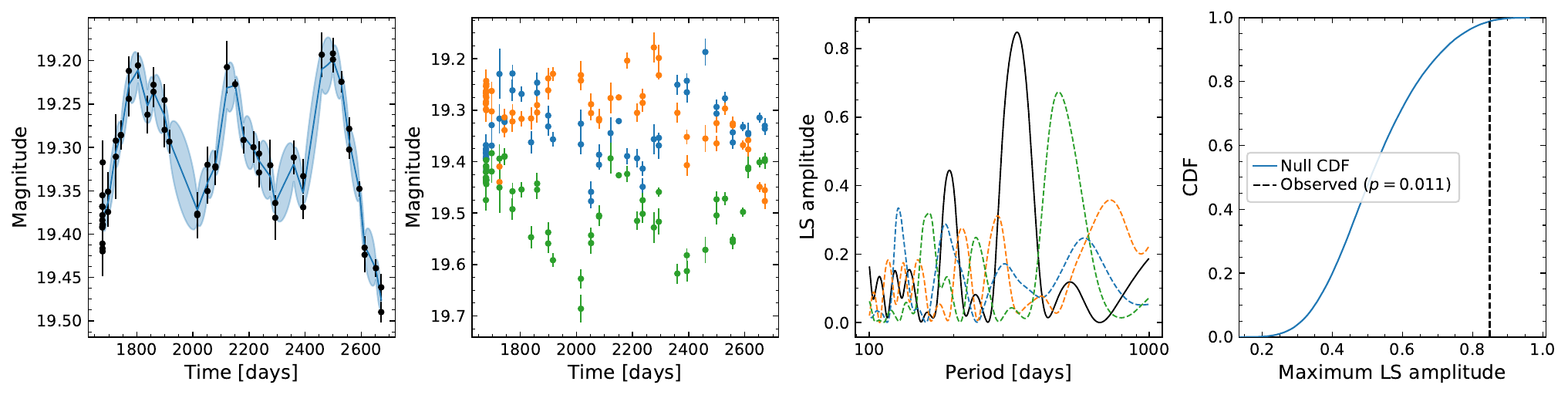}
    \caption{Illustration of the red-noise false-alarm probability estimation procedure using \gaia \texttt{4739629868656108160} as an example. The left panel shows the \g band photometry (black points) with the maximum
    a posteriori DRW GP fit overplotted. The second-most left panel shows three realizations drawn from the fitted parameters. The second-most right panel shows the LS periodogram of the observed data (solid black) and those of the three realizations (dashed colour). The right panel shows the empirical cumulative distribution function of the maximum LS amplitude across 100\,000 simulated time series. The black dashed vertical line marks the observed amplitude, corresponding to a $p$-value of $0.0111$, well above the adopted significance threshold of $\alpha=10^{-5}$.}
    \label{fig:methodology}
\end{figure*}

In this work, we assess the significance of periodogram peaks using 
empirical false alarm probabilities derived from DRW simulations \citep{charisi2016population, Chen2020, Chen+2024, Luo2025}. 
Figure~\ref{fig:methodology} summarises the steps in our pipeline. First, a red-noise model is fitted to the light curve of each source. The recovered posterior distribution is then used to draw realisations of synthetic red-noise light curves sampled at the same time stamps and with the same photometric uncertainties as the observations. The LS periodogram is computed for each realisation and the maximum amplitude is recorded. Finally, an empirical $p$-value is obtained as the fraction of synthetic maxima that exceed the observed maximum. The following subsections describe each of these steps in detail.

%%%%%%%%%%%%%%%%%%%%%%%%%%%%%%%%
\subsection{Red noise modelling}
%%%%%%%%%%%%%%%%%%%%%%%%%%%%%%%%
\label{sec:red_noise}

For a time series $\{t_i, m_i, e_i\}_{i=1,\ldots,N}$ we model each \gaia \g band light curve as a Gaussian process (GP) with constant mean $\mu$ and a stationary covariance kernel. GPs provide a flexible, non-parametric framework for regression and marginalisation over latent functions \citep{williams2006gaussian, aigrain2023gaussian}. Two kernel variants are considered. The first is the standard DRW (exponential) covariance,
\begin{equation}
  \kappa_\text{DRW}(t,t') = \sigma_R^2
    \exp\!\left(-\frac{|t-t'|}{\tau}\right),
\end{equation}
parametrised by a variance amplitude $\sigma_R^2$ and a characteristic timescale $\tau$. The second is the generalised powered-exponential covariance,
\begin{equation}
  \kappa_\gamma(t,t') = \sigma_R^2
    \exp\!\left(-\left(\frac{|t-t'|}{\tau}\right)^{\!\gamma}\right),
  \qquad \gamma\in[1,2],
\end{equation}
which reduces to the DRW kernel for $\gamma=1$. Allowing $\gamma>1$ produces steeper power spectral density slopes, enabling the model to account for AGN variability that is more correlated on short timescales than the DRW predicts. This more general kernel is applied only to the most significant candidates as an additional test of whether the detected periodicity survives under a less restrictive red-noise model.

We infer the model parameters within a Bayesian framework, placing the following priors:
\begin{align}
  \mu            &\sim \mathcal{N}\!\left(\bar{m},\,1\right), \\
  \log\sigma_R   &\sim \mathcal{N}\!\left(-2.3,\,1.15^2\right), \\
  \log\tau       &\sim \mathcal{N}\!\left(5.3,\,1.15^2\right), \\
  \operatorname{logit}\gamma &\sim \mathcal{N}(0,1),
\end{align}
where $\bar{m}$ is the sample mean of the observed magnitudes. Parameters $\sigma_R$ and $\tau$ are centred at 0.1 mag and 200 days, respectively \citep{macleod2010sdssdrw, vaughan2016false}. The exponent $\gamma$ is constrained to $[1,2]$ via a sigmoid reparametrisation of the unconstrained variable $\operatorname{logit}\gamma$.

Parameter inference proceeds in two stages. In the first stage, applied to all 377\,128 sources, the posterior is approximated via a Laplace approximation centred at the maximum \textit{a~posteriori} (MAP) estimate using the DRW kernel. Synthetic light curves for the Monte Carlo $p$-value estimation (Section~\ref{sec:mc_pval}) are generated by sampling parameters from this Gaussian approximation to the posterior and drawing realisations from the GP prior at each parameter sample. In the second stage, restricted to sources whose $p$-value falls below a threshold $\alpha$, the posterior is sampled directly using the No-U-Turn Sampler \citep[NUTS;][]{hoffman2011nuts}, run with four chains of 25\,000 samples each (after 500 warm-up iterations) and a target acceptance probability of 0.9. %Convergence is assessed using the $\hat{R}$ statistic \citep{gelman1992rubin}. 
In this stage, both kernel variants are tested.

The GP and kernel computations are implemented using \texttt{tinygp} \citep{foreman_mackey_2026_19035246} in JAX \citep{jax2018github}, and parameter inference is performed using the NumPyro probabilistic programming library \citep{phan2019numpyro}. This two-stage strategy makes the pipeline computationally tractable at the scale of the full sample while providing rigorous posterior characterisation for the most promising candidates.

%%%%%%%%%%%%%%%%%%%%%%%%%%%%%%%%%%%%
\subsection{Periodogram computation}
%%%%%%%%%%%%%%%%%%%%%%%%%%%%%%%%%%%%

The LS periodogram is computed for each light curve and for every synthetic red-noise realisation. For a light curve of duration $T = \max(t) - \min(t)$, the frequency grid spans  
$f_\text{min}=1/T~\text{d}^{-1}$ to $f_\text{max}=0.01~\text{d}^{-1}$ with a uniform step of $\Delta_f = (100 T)^{-1}~\text{d}^{-1}$. 
This choice of $f_\text{min}$ corresponds to a period of 950 \,d on average, while the value of
$f_\text{max}$ avoids the majority of spurious periodicities reported by \citet{holl2023gaia}. Moreover, according to \citet{park2024self}, the likelihood of detecting genuine SMBHB candidates at periods shorter than 100\,d is exceedingly low. The periodogram is evaluated efficiently using the NUFFT-based implementation of \citet{garrison2024nifty}. For reference, the analytical FAP of \citet{baluev2008fap}, as implemented in \texttt{astropy} \citep{astropy}, is also recorded.

%%%%%%%%%%%%%%%%%%%%%%%%%%%%%%%%%%%%%%%%%%%%%
\subsection{Monte Carlo $p$-value estimation}
%%%%%%%%%%%%%%%%%%%%%%%%%%%%%%%%%%%%%%%%%%%%%
\label{sec:mc_pval}

For each source, $N_\text{sim} = 100\,000$ synthetic light curves are
generated by sampling the GP at the observed time stamps and with
the observed photometric uncertainties, using parameter values drawn
from the approximate posterior in the first stage or the MCMC samples
directly in the second stage.

The empirical $p$-value is estimated as
\begin{equation}
  \hat{p} = \frac{C + \tfrac{1}{2}}{N_\text{sim} + 1},
\end{equation}
where $C$ is the number of simulations whose maximum amplitude exceeds the observed peak amplitude. This smoothed estimator corresponds to a Jeffreys (Beta$(\tfrac{1}{2},\tfrac{1}{2})$) prior on the true $p$-value and avoids $\hat{p}_k = 0$ for non-detections. 

%%%%%%%%%%%%%%%%%%%%%%%%%%%%%%%%%%%%%%%%%%%%%%%%%
\section{Analysis results}
\label{sec:results}
%%%%%%%%%%%%%%%%%%%%%%%%%%%%%%%%%%%%%%%%%%%%%%%%%
 
\begin{figure}[t]
    \centering
    \includegraphics[width=0.48\textwidth]{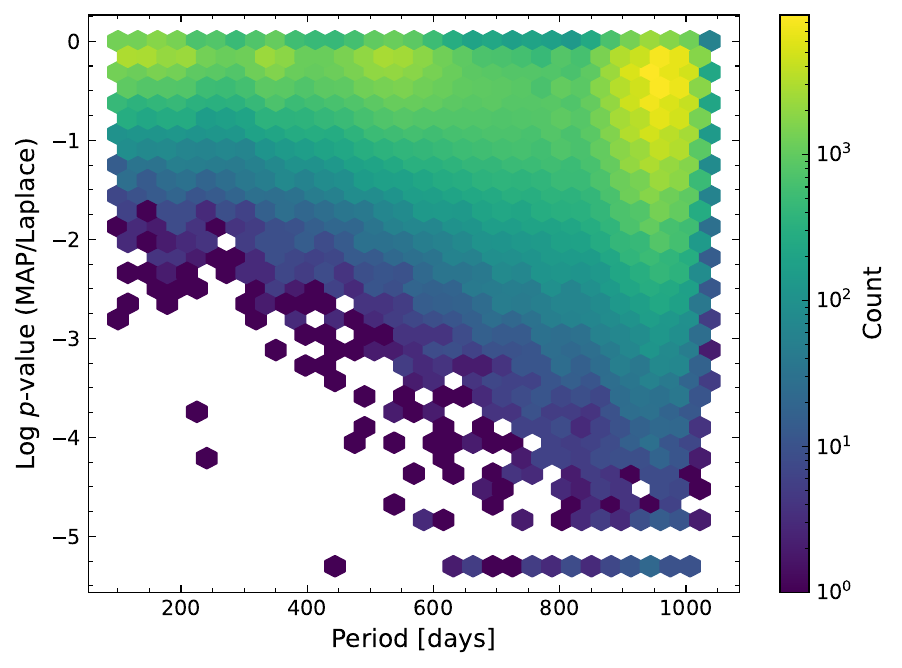}
    \caption{Distribution of dominant period against Monte Carlo $p$-value from the MAP/Laplace filter for the full sample. }
    \label{fig:period_vs_fap}
\end{figure}

The analytical FAP of \citet{baluev2008fap} flags 305\,116 of the 377\,119 sources (80.9\%) as having significant periodogram peaks, illustrating the inadequacy of a white-noise null hypothesis for this population. Under our red-noise calibrated test, 53 sources survive the MAP and Laplace approximation filter (first stage) with $p < \alpha = 10^{-5}$. At this significance threshold, the expected number of false positives is $\sim$3.8, so the 53 detections represent a modest excess above the chance expectation. Figure~\ref{fig:period_vs_fap} shows the distribution of dominant period against the first-stage $p$-value for all sources. The most significant detections are overwhelmingly concentrated at periods approaching the observational baseline. 

Requiring each candidate to additionally pass both MCMC filters, for both the exponential and the powered-exponential kernel, reduces the sample to 13 sources (Table~\ref{tab:candidates}). Of these, only one covers more than 1.5 cycles\footnote{The minimum cycle requirements adopted in previous searches \citep{graham2015systematic, charisi2016population}.}: a best-fit period of ${\sim}459$\,days corresponding to approximately 2.1 observed cycles (Figure~\ref{fig:best_candidates}, top panel). Even the single candidate at ${\sim}2.1$ cycles falls within the 1.5--2.5 cycle regime for which \citet{vaughan2016false} demonstrate that phantom periodicities readily emerge from steep-spectrum stochastic processes.

\begin{figure}[t]
    \centering
    \includegraphics[width=0.48\textwidth]{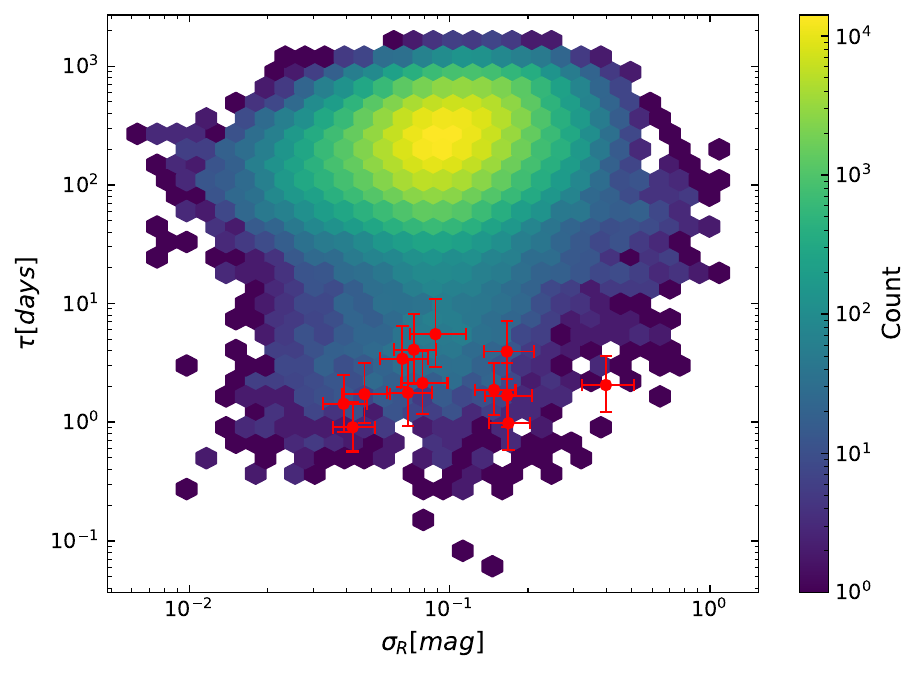}
    \caption{Distribution of fitted DRW parameters $\sigma_R$ and $\tau$ for the full sample. The 13 candidates surviving both MCMC filters are indicated as red dots with error bars.}
    \label{fig:tau_sigma}
\end{figure}

Figure~\ref{fig:tau_sigma} reveals that the 13 surviving candidates are systematically concentrated in the low-$\tau$ tail of the DRW parameter distribution, with characteristic timescales of order 1--10\,days. These sources exhibit coherent short-timescale variability resolved within \gaia's sampling pattern alongside smoother long-timescale trends. A single-DRW kernel cannot capture both components simultaneously. Rather, it fits the short timescale, leaving the long-timescale variability unmodelled and therefore unaccounted for in the null distribution. The low $p$-values of these candidates therefore likely reflect model misspecification rather than genuine periodicity.

%%%%%%%%%%%%%%%%%%%%%%%%%%%%%%%%%%%%%%%%%%%%%%%
\section{Conclusions}
%%%%%%%%%%%%%%%%%%%%%%%%%%%%%%%%%%%%%%%%%%%%%%%
\label{sec:conclusions}

We have designed an analysis pipeline to
search for periodic variability in \gaia DR3 light curves of AGN and applied it to
377\,128 sources. We calibrated the periodogram significance against red-noise null models with progressively more rigorous posterior inference. No reliable periodic candidates are detected. The 13 candidates surviving the most stringent filters are consistent with the expected false-positive rate, cover fewer than 2.5 observed cycles, and occupy a region of DRW parameter space indicative of null model misspecification.

\gaia's photometric precision surpasses that of ground-based surveys \citep[see figures in][]{El-Badry_2026}. Yet the ${\sim}950$-day baseline of DR3 confines all detections to the few-cycle regime where stochastic variability most convincingly mimics periodicity \citep{vaughan2016false}. Our current sample is further shaped by the DR3 release strategy, which restricts epoch photometry to sources flagged as photometrically variable, a subset biased toward high variability amplitudes. 
The complete epoch photometry expected in DR4, combined with its longer baseline, will enable an unbiased search across the full AGN population and push detections into the multi-cycle regime where periodic and stochastic signals can be more reliably distinguished. Our developed analysis pipeline\footnote{\url{https://github.com/IvS-KULeuven/GASP}} is readily applicable to efficiently exploit the \gaia DR4 data to be released in Dec.\,2026.

\begin{acknowledgements} 
This work has made use of data from the European Space Agency (ESA) mission \gaia (\url{https://www.cosmos.esa.int/gaia}), processed by the \gaia Data Processing and Analysis Consortium (DPAC,
\url{https://www.cosmos.esa.int/web/gaia/dpac/consortium}). Funding for the DPAC has been provided by national institutions, in particular the institutions participating in the \gaia Multilateral Agreement. 
PH, JDR and CA acknowledge support from the BELgian federal Science Policy Office (BELSPO) through various PROgramme de Développement d’EXpériences scientifiques (PRODEX) grants to develop \gaia variability data analysis pipelines.
%PH and CA acknowledge financial support from the European Research Council (ERC) under the Horizon Europe programme (Synergy Grant agreement N◦101071505: 4D-STAR). 
PH also acknowledges support from ANID – Millennium Science Initiative Program – ICN12\_009 awarded to the Millennium Institute of Astrophysics MAS. 
This research was supported by the Patag\'{o}n supercomputer of Universidad Austral de Chile (FONDEQUIP EQM180042).
We acknowledge the fruitful constructive suggestions that we received from the community following our original draft manuscript placed on arXiv. In particular, we thank Di Luo, Kareem El-Badry, and anonymous reviewers whose comments and suggestions helped shape the methodology presented in this work.
\end{acknowledgements}

\bibliographystyle{aa} 
\bibliography{references}

@article{vallenari2023gaia,
       author = {{Gaia Collaboration} and {Vallenari}, A. and {Brown}, A.~G.~A. and {Prusti}, T. and {de Bruijne}, J.~H.~J. and {Arenou}, F. and {Babusiaux}, C. and {Biermann}, M. and {Creevey}, O.~L. and {Ducourant}, C. and {Evans}, D.~W. and {Eyer}, L. and {Guerra}, R. and {Hutton}, A. and {Jordi}, C. and {Klioner}, S.~A. and {Lammers}, U.~L. and {Lindegren}, L. and {Luri}, X. and {Mignard}, F. and {Panem}, C. and {Pourbaix}, D. and {Randich}, S. and {Sartoretti}, P. and {Soubiran}, C. and {Tanga}, P. and {Walton}, N.~A. and {Bailer-Jones}, C.~A.~L. and {Bastian}, U. and {Drimmel}, R. and {Jansen}, F. and {Katz}, D. and {Lattanzi}, M.~G. and {van Leeuwen}, F. and {Bakker}, J. and {Cacciari}, C. and {Casta{\~n}eda}, J. and {De Angeli}, F. and {Fabricius}, C. and {Fouesneau}, M. and {Fr{\'e}mat}, Y. and {Galluccio}, L. and {Guerrier}, A. and {Heiter}, U. and {Masana}, E. and {Messineo}, R. and {Mowlavi}, N. and {Nicolas}, C. and {Nienartowicz}, K. and {Pailler}, F. and {Panuzzo}, P. and {Riclet}, F. and {Roux}, W. and {Seabroke}, G.~M. and {Sordo}, R. and {Th{\'e}venin}, F. and {Gracia-Abril}, G. and {Portell}, J. and {Teyssier}, D. and {Altmann}, M. and {Andrae}, R. and {Audard}, M. and {Bellas-Velidis}, I. and {Benson}, K. and {Berthier}, J. and {Blomme}, R. and {Burgess}, P.~W. and {Busonero}, D. and {Busso}, G. and {C{\'a}novas}, H. and {Carry}, B. and {Cellino}, A. and {Cheek}, N. and {Clementini}, G. and {Damerdji}, Y. and {Davidson}, M. and {de Teodoro}, P. and {Nu{\~n}ez Campos}, M. and {Delchambre}, L. and {Dell'Oro}, A. and {Esquej}, P. and {Fern{\'a}ndez-Hern{\'a}ndez}, J. and {Fraile}, E. and {Garabato}, D. and {Garc{\'\i}a-Lario}, P. and {Gosset}, E. and {Haigron}, R. and {Halbwachs}, J. -L. and {Hambly}, N.~C. and {Harrison}, D.~L. and {Hern{\'a}ndez}, J. and {Hestroffer}, D. and {Hodgkin}, S.~T. and {Holl}, B. and {Jan{\ss}en}, K. and {Jevardat de Fombelle}, G. and {Jordan}, S. and {Krone-Martins}, A. and {Lanzafame}, A.~C. and {L{\"o}ffler}, W. and {Marchal}, O. and {Marrese}, P.~M. and {Moitinho}, A. and {Muinonen}, K. and {Osborne}, P. and {Pancino}, E. and {Pauwels}, T. and {Recio-Blanco}, A. and {Reyl{\'e}}, C. and {Riello}, M. and {Rimoldini}, L. and {Roegiers}, T. and {Rybizki}, J. and {Sarro}, L.~M. and {Siopis}, C. and {Smith}, M. and {Sozzetti}, A. and {Utrilla}, E. and {van Leeuwen}, M. and {Abbas}, U. and {{\'A}brah{\'a}m}, P. and {Abreu Aramburu}, A. and {Aerts}, C. and {Aguado}, J.~J. and {Ajaj}, M. and {Aldea-Montero}, F. and {Altavilla}, G. and {{\'A}lvarez}, M.~A. and {Alves}, J. and {Anders}, F. and {Anderson}, R.~I. and {Anglada Varela}, E. and {Antoja}, T. and {Baines}, D. and {Baker}, S.~G. and {Balaguer-N{\'u}{\~n}ez}, L. and {Balbinot}, E. and {Balog}, Z. and {Barache}, C. and {Barbato}, D. and {Barros}, M. and {Barstow}, M.~A. and {Bartolom{\'e}}, S. and {Bassilana}, J. -L. and {Bauchet}, N. and {Becciani}, U. and {Bellazzini}, M. and {Berihuete}, A. and {Bernet}, M. and {Bertone}, S. and {Bianchi}, L. and {Binnenfeld}, A. and {Blanco-Cuaresma}, S. and {Blazere}, A. and {Boch}, T. and {Bombrun}, A. and {Bossini}, D. and {Bouquillon}, S. and {Bragaglia}, A. and {Bramante}, L. and {Breedt}, E. and {Bressan}, A. and {Brouillet}, N. and {Brugaletta}, E. and {Bucciarelli}, B. and {Burlacu}, A. and {Butkevich}, A.~G. and {Buzzi}, R. and {Caffau}, E. and {Cancelliere}, R. and {Cantat-Gaudin}, T. and {Carballo}, R. and {Carlucci}, T. and {Carnerero}, M.~I. and {Carrasco}, J.~M. and {Casamiquela}, L. and {Castellani}, M. and {Castro-Ginard}, A. and {Chaoul}, L. and {Charlot}, P. and {Chemin}, L. and {Chiaramida}, V. and {Chiavassa}, A. and {Chornay}, N. and {Comoretto}, G. and {Contursi}, G. and {Cooper}, W.~J. and {Cornez}, T. and {Cowell}, S. and {Crifo}, F. and {Cropper}, M. and {Crosta}, M. and {Crowley}, C. and {Dafonte}, C. and {Dapergolas}, A. and {David}, M. and {David}, P. and {de Laverny}, P. and {De Luise}, F. and {De March}, R.},
        title = "{Gaia Data Release 3. Summary of the content and survey properties}",
      journal = {\aap},
         year = 2023,
        month = jun,
       volume = {674},
          eid = {A1},
        pages = {A1},
          doi = {10.1051/0004-6361/202243940},
archivePrefix = {arXiv},
       eprint = {2208.00211},
 primaryClass = {astro-ph.GA},
       adsurl = {https://ui.adsabs.harvard.edu/abs/2023A&A...674A...1G}
}

@article{klioner2022gaia,
       author = {{Gaia Collaboration} and {Klioner}, S.~A. and {Lindegren}, L. and {Mignard}, F. and {Hern{\'a}ndez}, J. and {Ramos-Lerate}, M. and {Bastian}, U. and {Biermann}, M. and {Bombrun}, A. and {de Torres}, A. and {Gerlach}, E. and {Geyer}, R. and {Hilger}, T. and {Hobbs}, D. and {Lammers}, U.~L. and {McMillan}, P.~J. and {Steidelm{\"u}ller}, H. and {Teyssier}, D. and {Raiteri}, C.~M. and {Bartolom{\'e}}, S. and {Bernet}, M. and {Casta{\~n}eda}, J. and {Clotet}, M. and {Davidson}, M. and {Fabricius}, C. and {Garralda Torres}, N. and {Gonz{\'a}lez-Vidal}, J.~J. and {Portell}, J. and {Rowell}, N. and {Torra}, F. and {Torra}, J. and {Brown}, A.~G.~A. and {Vallenari}, A. and {Prusti}, T. and {de Bruijne}, J.~H.~J. and {Arenou}, F. and {Babusiaux}, C. and {Creevey}, O.~L. and {Ducourant}, C. and {Evans}, D.~W. and {Eyer}, L. and {Guerra}, R. and {Hutton}, A. and {Jordi}, C. and {Luri}, X. and {Panem}, C. and {Pourbaix}, D. and {Randich}, S. and {Sartoretti}, P. and {Soubiran}, C. and {Tanga}, P. and {Walton}, N.~A. and {Bailer-Jones}, C.~A.~L. and {Drimmel}, R. and {Jansen}, F. and {Katz}, D. and {Lattanzi}, M.~G. and {van Leeuwen}, F. and {Bakker}, J. and {Cacciari}, C. and {De Angeli}, F. and {Fouesneau}, M. and {Fr{\'e}mat}, Y. and {Galluccio}, L. and {Guerrier}, A. and {Heiter}, U. and {Masana}, E. and {Messineo}, R. and {Mowlavi}, N. and {Nicolas}, C. and {Nienartowicz}, K. and {Pailler}, F. and {Panuzzo}, P. and {Riclet}, F. and {Roux}, W. and {Seabroke}, G.~M. and {Sordo}, R. and {Th{\'e}venin}, F. and {Gracia-Abril}, G. and {Altmann}, M. and {Andrae}, R. and {Audard}, M. and {Bellas-Velidis}, I. and {Benson}, K. and {Berthier}, J. and {Blomme}, R. and {Burgess}, P.~W. and {Busonero}, D. and {Busso}, G. and {C{\'a}novas}, H. and {Carry}, B. and {Cellino}, A. and {Cheek}, N. and {Clementini}, G. and {Damerdji}, Y. and {de Teodoro}, P. and {Nu{\~n}ez Campos}, M. and {Delchambre}, L. and {Dell'Oro}, A. and {Esquej}, P. and {Fern{\'a}ndez-Hern{\'a}ndez}, J. and {Fraile}, E. and {Garabato}, D. and {Garc{\'\i}a-Lario}, P. and {Gosset}, E. and {Haigron}, R. and {Halbwachs}, J. -L. and {Hambly}, N.~C. and {Harrison}, D.~L. and {Hestroffer}, D. and {Hodgkin}, S.~T. and {Holl}, B. and {Jan{\ss}en}, K. and {Jevardat de Fombelle}, G. and {Jordan}, S. and {Krone-Martins}, A. and {Lanzafame}, A.~C. and {L{\"o}ffler}, W. and {Marchal}, O. and {Marrese}, P.~M. and {Moitinho}, A. and {Muinonen}, K. and {Osborne}, P. and {Pancino}, E. and {Pauwels}, T. and {Recio-Blanco}, A. and {Reyl{\'e}}, C. and {Riello}, M. and {Rimoldini}, L. and {Roegiers}, T. and {Rybizki}, J. and {Sarro}, L.~M. and {Siopis}, C. and {Smith}, M. and {Sozzetti}, A. and {Utrilla}, E. and {van Leeuwen}, M. and {Abbas}, U. and {{\'A}brah{\'a}m}, P. and {Abreu Aramburu}, A. and {Aerts}, C. and {Aguado}, J.~J. and {Ajaj}, M. and {Aldea-Montero}, F. and {Altavilla}, G. and {{\'A}lvarez}, M.~A. and {Alves}, J. and {Anderson}, R.~I. and {Anglada Varela}, E. and {Antoja}, T. and {Baines}, D. and {Baker}, S.~G. and {Balaguer-N{\'u}{\~n}ez}, L. and {Balbinot}, E. and {Balog}, Z. and {Barache}, C. and {Barbato}, D. and {Barros}, M. and {Barstow}, M.~A. and {Bassilana}, J. -L. and {Bauchet}, N. and {Becciani}, U. and {Bellazzini}, M. and {Berihuete}, A. and {Bertone}, S. and {Bianchi}, L. and {Binnenfeld}, A. and {Blanco-Cuaresma}, S. and {Boch}, T. and {Bossini}, D. and {Bouquillon}, S. and {Bragaglia}, A. and {Bramante}, L. and {Breedt}, E. and {Bressan}, A. and {Brouillet}, N. and {Brugaletta}, E. and {Bucciarelli}, B. and {Burlacu}, A. and {Butkevich}, A.~G. and {Buzzi}, R. and {Caffau}, E. and {Cancelliere}, R. and {Cantat-Gaudin}, T. and {Carballo}, R. and {Carlucci}, T. and {Carnerero}, M.~I. and {Carrasco}, J.~M. and {Casamiquela}, L. and {Castellani}, M. and {Castro-Ginard}, A. and {Chaoul}, L. and {Charlot}, P. and {Chemin}, L. and {Chiaramida}, V. and {Chiavassa}, A. and {Chornay}, N. and {Comoretto}, G. and {Contursi}, G. and {Cooper}, W.~J.},
        title = "{Gaia Early Data Release 3. The celestial reference frame (Gaia-CRF3)}",
      journal = {\aap},
         year = 2022,
        month = nov,
       volume = {667},
          eid = {A148},
        pages = {A148},
          doi = {10.1051/0004-6361/202243483},
archivePrefix = {arXiv},
       eprint = {2204.12574},
 primaryClass = {astro-ph.IM},
       adsurl = {https://ui.adsabs.harvard.edu/abs/2022A&A...667A.148G}
}

@article{bailer2023gaia,
       author = {{Gaia Collaboration} and {Bailer-Jones}, C.~A.~L. and {Teyssier}, D. and {Delchambre}, L. and {Ducourant}, C. and {Garabato}, D. and {Hatzidimitriou}, D. and {Klioner}, S.~A. and {Rimoldini}, L. and {Bellas-Velidis}, I. and {Carballo}, R. and {Carnerero}, M.~I. and {Diener}, C. and {Fouesneau}, M. and {Galluccio}, L. and {Gavras}, P. and {Krone-Martins}, A. and {Raiteri}, C.~M. and {Teixeira}, R. and {Brown}, A.~G.~A. and {Vallenari}, A. and {Prusti}, T. and {de Bruijne}, J.~H.~J. and {Arenou}, F. and {Babusiaux}, C. and {Biermann}, M. and {Creevey}, O.~L. and {Evans}, D.~W. and {Eyer}, L. and {Guerra}, R. and {Hutton}, A. and {Jordi}, C. and {Lammers}, U.~L. and {Lindegren}, L. and {Luri}, X. and {Mignard}, F. and {Panem}, C. and {Pourbaix}, D. and {Randich}, S. and {Sartoretti}, P. and {Soubiran}, C. and {Tanga}, P. and {Walton}, N.~A. and {Bastian}, U. and {Drimmel}, R. and {Jansen}, F. and {Katz}, D. and {Lattanzi}, M.~G. and {van Leeuwen}, F. and {Bakker}, J. and {Cacciari}, C. and {Casta{\~n}eda}, J. and {De Angeli}, F. and {Fabricius}, C. and {Fr{\'e}mat}, Y. and {Guerrier}, A. and {Heiter}, U. and {Masana}, E. and {Messineo}, R. and {Mowlavi}, N. and {Nicolas}, C. and {Nienartowicz}, K. and {Pailler}, F. and {Panuzzo}, P. and {Riclet}, F. and {Roux}, W. and {Seabroke}, G.~M. and {Sordo}, R. and {Th{\'e}venin}, F. and {Gracia-Abril}, G. and {Portell}, J. and {Altmann}, M. and {Andrae}, R. and {Audard}, M. and {Benson}, K. and {Berthier}, J. and {Blomme}, R. and {Burgess}, P.~W. and {Busonero}, D. and {Busso}, G. and {C{\'a}novas}, H. and {Carry}, B. and {Cellino}, A. and {Cheek}, N. and {Clementini}, G. and {Damerdji}, Y. and {Davidson}, M. and {de Teodoro}, P. and {Nu{\~n}ez Campos}, M. and {Dell'Oro}, A. and {Esquej}, P. and {Fern{\'a}ndez-Hern{\'a}ndez}, J. and {Fraile}, E. and {Garc{\'\i}a-Lario}, P. and {Gosset}, E. and {Haigron}, R. and {Halbwachs}, J. -L. and {Hambly}, N.~C. and {Harrison}, D.~L. and {Hern{\'a}ndez}, J. and {Hestroffer}, D. and {Hodgkin}, S.~T. and {Holl}, B. and {Jan{\ss}en}, K. and {Jevardat de Fombelle}, G. and {Jordan}, S. and {Lanzafame}, A.~C. and {L{\"o}ffler}, W. and {Marchal}, O. and {Marrese}, P.~M. and {Moitinho}, A. and {Muinonen}, K. and {Osborne}, P. and {Pancino}, E. and {Pauwels}, T. and {Recio-Blanco}, A. and {Reyl{\'e}}, C. and {Riello}, M. and {Roegiers}, T. and {Rybizki}, J. and {Sarro}, L.~M. and {Siopis}, C. and {Smith}, M. and {Sozzetti}, A. and {Utrilla}, E. and {van Leeuwen}, M. and {Abbas}, U. and {{\'A}brah{\'a}m}, P. and {Abreu Aramburu}, A. and {Aerts}, C. and {Aguado}, J.~J. and {Ajaj}, M. and {Aldea-Montero}, F. and {Altavilla}, G. and {{\'A}lvarez}, M.~A. and {Alves}, J. and {Anderson}, R.~I. and {Anglada Varela}, E. and {Antoja}, T. and {Baines}, D. and {Baker}, S.~G. and {Balaguer-N{\'u}{\~n}ez}, L. and {Balbinot}, E. and {Balog}, Z. and {Barache}, C. and {Barbato}, D. and {Barros}, M. and {Barstow}, M.~A. and {Bartolom{\'e}}, S. and {Bassilana}, J. -L. and {Bauchet}, N. and {Becciani}, U. and {Bellazzini}, M. and {Berihuete}, A. and {Bernet}, M. and {Bertone}, S. and {Bianchi}, L. and {Binnenfeld}, A. and {Blanco-Cuaresma}, S. and {Boch}, T. and {Bombrun}, A. and {Bossini}, D. and {Bouquillon}, S. and {Bragaglia}, A. and {Bramante}, L. and {Breedt}, E. and {Bressan}, A. and {Brouillet}, N. and {Brugaletta}, E. and {Bucciarelli}, B. and {Burlacu}, A. and {Butkevich}, A.~G. and {Buzzi}, R. and {Caffau}, E. and {Cancelliere}, R. and {Cantat-Gaudin}, T. and {Carlucci}, T. and {Carrasco}, J.~M. and {Casamiquela}, L. and {Castellani}, M. and {Castro-Ginard}, A. and {Chaoul}, L. and {Charlot}, P. and {Chemin}, L. and {Chiaramida}, V. and {Chiavassa}, A. and {Chornay}, N. and {Comoretto}, G. and {Contursi}, G. and {Cooper}, W.~J. and {Cornez}, T. and {Cowell}, S. and {Crifo}, F. and {Cropper}, M. and {Crosta}, M. and {Crowley}, C. and {Dafonte}, C. and {Dapergolas}, A. and {David}, P. and {de Laverny}, P.},
        title = "{Gaia Data Release 3. The extragalactic content}",
      journal = {\aap},
         year = 2023,
        month = jun,
       volume = {674},
          eid = {A41},
        pages = {A41},
          doi = {10.1051/0004-6361/202243232},
archivePrefix = {arXiv},
       eprint = {2206.05681},
 primaryClass = {astro-ph.GA},
       adsurl = {https://ui.adsabs.harvard.edu/abs/2023A&A...674A..41G}
}

@ARTICLE{carnenero2023gaia,
       author = {{Carnerero}, Maria I. and {Raiteri}, Claudia M. and {Rimoldini}, Lorenzo and {Busonero}, Deborah and {Licata}, Enrico and {Mowlavi}, Nami and {Lecoeur-Ta{\"\i}bi}, Isabelle and {Audard}, Marc and {Holl}, Berry and {Gavras}, Panagiotis and {Nienartowicz}, Krzysztof and {Jevardat de Fombelle}, Gr{\'e}gory and {Carballo}, Ruth and {Clementini}, Gisella and {Delchambre}, Ludovic and {Klioner}, Sergei and {Lattanzi}, Mario G. and {Eyer}, Laurent},
        title = "{Gaia Data Release 3. The first Gaia catalogue of variable AGN}",
      journal = {\aap},
         year = 2023,
        month = jun,
       volume = {674},
          eid = {A24},
        pages = {A24},
          doi = {10.1051/0004-6361/202244035},
archivePrefix = {arXiv},
       eprint = {2207.06849},
 primaryClass = {astro-ph.HE},
       adsurl = {https://ui.adsabs.harvard.edu/abs/2023A&A...674A..24C}
}

@ARTICLE{holl2023gaia,
       author = {{Holl}, B. and {Fabricius}, C. and {Portell}, J. and {Lindegren}, L. and {Panuzzo}, P. and {Bernet}, M. and {Casta{\~n}eda}, J. and {Jevardat de Fombelle}, G. and {Audard}, M. and {Ducourant}, C. and {Harrison}, D.~L. and {Evans}, D.~W. and {Busso}, G. and {Sozzetti}, A. and {Gosset}, E. and {Arenou}, F. and {De Angeli}, F. and {Riello}, M. and {Eyer}, L. and {Rimoldini}, L. and {Gavras}, P. and {Mowlavi}, N. and {Nienartowicz}, K. and {Lecoeur-Ta{\"\i}bi}, I. and {Garc{\'\i}a-Lario}, P. and {Pourbaix}, D.},
        title = "{Gaia Data Release 3. Gaia scan-angle-dependent signals and spurious periods}",
      journal = {\aap},
         year = 2023,
        month = jun,
       volume = {674},
          eid = {A25},
        pages = {A25},
          doi = {10.1051/0004-6361/202245353},
archivePrefix = {arXiv},
       eprint = {2212.11971},
 primaryClass = {astro-ph.IM},
       adsurl = {https://ui.adsabs.harvard.edu/abs/2023A&A...674A..25H}
}

@MISC{rimoldini2022gaiadr3manual,
       author = {{Rimoldini}, L. and {Eyer}, L. and {Audard}, M. and {Barblan}, F. and {Carnerero}, M.~I. and {Clementini}, G. and {De Ridder}, J. and {Distefano}, E. and {Faigler}, S. and {Garofalo}, A. and {Gavras}, P. and {Gomel}, R. and {Holl}, B. and {Jevardat de Fombelle}, G. and {Kruszy{\'n}ska}, K. and {Lanzafame}, A. and {Lebzelter}, T. and {Leccia}, S. and {Lecoeur-Ta{\"\i}bi}, I. and {Mazeh}, T. and {Molinaro}, R. and {Mowlavi}, N. and {Muraveva}, T. and {Nienartowicz}, K. and {Panahi}, A. and {Raiteri}, C.~M. and {Ripepi}, V. and {Rybicki}, K.~A. and {Trabucchi}, M. and {Wyrzykowski}, {\L}. and {Zucker}, S.},
        title = "{Gaia DR3 documentation Chapter 10: Variability}",
 howpublished = {Gaia DR3 documentation, European Space Agency; Gaia Data Processing and Analysis Consortium. Online at \url{https://gea.esac.esa.int/archive/documentation/GDR3/index.html}, id. 10},
         year = 2022,
        month = jun,
          eid = {10},
        pages = {10},
       adsurl = {https://ui.adsabs.harvard.edu/abs/2022gdr3.reptE..10R}
}

@ARTICLE{begelman1980smbhb,
       author = {{Begelman}, M.~C. and {Blandford}, R.~D. and {Rees}, M.~J.},
        title = "{Massive black hole binaries in active galactic nuclei}",
      journal = {\nat},
         year = 1980,
        month = sep,
       volume = {287},
       number = {5780},
        pages = {307-309},
          doi = {10.1038/287307a0},
       adsurl = {https://ui.adsabs.harvard.edu/abs/1980Natur.287..307B}
}

@ARTICLE{dorazio2013accretion,
       author = {{D'Orazio}, Daniel J. and {Haiman}, Zolt{\'a}n and {MacFadyen}, Andrew},
        title = "{Accretion into the central cavity of a circumbinary disc}",
      journal = {\mnras},
         year = 2013,
        month = dec,
       volume = {436},
       number = {4},
        pages = {2997-3020},
          doi = {10.1093/mnras/stt1787},
archivePrefix = {arXiv},
       eprint = {1210.0536},
 primaryClass = {astro-ph.GA},
       adsurl = {https://ui.adsabs.harvard.edu/abs/2013MNRAS.436.2997D}
}

@ARTICLE{shi2012accretion,
       author = {{Shi}, Ji-Ming and {Krolik}, Julian H. and {Lubow}, Stephen H. and {Hawley}, John F.},
        title = "{Three-dimensional Magnetohydrodynamic Simulations of Circumbinary Accretion Disks: Disk Structures and Angular Momentum Transport}",
      journal = {\apj},
         year = 2012,
        month = apr,
       volume = {749},
       number = {2},
          eid = {118},
        pages = {118},
          doi = {10.1088/0004-637X/749/2/118},
archivePrefix = {arXiv},
       eprint = {1110.4866},
 primaryClass = {astro-ph.HE},
       adsurl = {https://ui.adsabs.harvard.edu/abs/2012ApJ...749..118S}
}

@ARTICLE{dorazio2015dopplerboost,
       author = {{D'Orazio}, Daniel J. and {Haiman}, Zolt{\'a}n and {Schiminovich}, David},
        title = "{Relativistic boost as the cause of periodicity in a massive black-hole binary candidate}",
      journal = {\nat},
         year = 2015,
        month = sep,
       volume = {525},
       number = {7569},
        pages = {351-353},
          doi = {10.1038/nature15262},
archivePrefix = {arXiv},
       eprint = {1509.04301},
 primaryClass = {astro-ph.HE},
       adsurl = {https://ui.adsabs.harvard.edu/abs/2015Natur.525..351D}
}

@article{davelaar2022,
	title = {Self-Lensing Flares from Black Hole Binaries: {{General-relativistic}} Ray Tracing of Black Hole Binaries},
	shorttitle = {Self-Lensing Flares from Black Hole Binaries},
	author = {Davelaar, Jordy and Haiman, Zolt{\'a}n},
	year = {2022},
	month = {may},
	journal = {\prd},
	volume = {105},
	number = {10},
	pages = {103010},
	publisher = {{American Physical Society}},
	doi = {10.1103/PhysRevD.105.103010},
	urldate = {2023-02-22}
}

@ARTICLE{Hu2020,
       author = {{Hu}, Betty X. and {D'Orazio}, Daniel J. and {Haiman}, Zolt{\'a}n and {Smith}, Krista Lynne and {Snios}, Bradford and {Charisi}, Maria and {Di Stefano}, Rosanne},
        title = "{Spikey: self-lensing flares from eccentric SMBH binaries}",
      journal = {\mnras},
         year = 2020,
        month = jul,
       volume = {495},
       number = {4},
        pages = {4061-4070},
          doi = {10.1093/mnras/staa1312},
archivePrefix = {arXiv},
       eprint = {1910.05348},
 primaryClass = {astro-ph.HE},
       adsurl = {https://ui.adsabs.harvard.edu/abs/2020MNRAS.495.4061H}
}

@article{kelly2009variations,
       author = {{Kelly}, Brandon C. and {Bechtold}, Jill and {Siemiginowska}, Aneta},
        title = "{Are the Variations in Quasar Optical Flux Driven by Thermal Fluctuations?}",
      journal = {\apj},
         year = 2009,
        month = jun,
       volume = {698},
       number = {1},
        pages = {895-910},
          doi = {10.1088/0004-637X/698/1/895},
archivePrefix = {arXiv},
       eprint = {0903.5315},
 primaryClass = {astro-ph.CO},
       adsurl = {https://ui.adsabs.harvard.edu/abs/2009ApJ...698..895K}
}

@article{kozlowski2009quantifying,
       author = {{Koz{\l}owski}, Szymon and {Kochanek}, Christopher S. and {Udalski}, A. and {Wyrzykowski}, {\L}. and {Soszy{\'n}ski}, I. and {Szyma{\'n}ski}, M.~K. and {Kubiak}, M. and {Pietrzy{\'n}ski}, G. and {Szewczyk}, O. and {Ulaczyk}, K. and {Poleski}, R. and {OGLE Collaboration}},
        title = "{Quantifying Quasar Variability as Part of a General Approach to Classifying Continuously Varying Sources}",
      journal = {\apj},
         year = 2010,
        month = jan,
       volume = {708},
       number = {2},
        pages = {927-945},
          doi = {10.1088/0004-637X/708/2/927},
archivePrefix = {arXiv},
       eprint = {0909.1326},
 primaryClass = {astro-ph.CO},
       adsurl = {https://ui.adsabs.harvard.edu/abs/2010ApJ...708..927K}
}

@ARTICLE{macleod2010sdssdrw,
       author = {{MacLeod}, C.~L. and {Ivezi{\'c}}, {\v{Z}}. and {Kochanek}, C.~S. and {Koz{\l}owski}, S. and {Kelly}, B. and {Bullock}, E. and {Kimball}, A. and {Sesar}, B. and {Westman}, D. and {Brooks}, K. and {Gibson}, R. and {Becker}, A.~C. and {de Vries}, W.~H.},
        title = "{Modeling the Time Variability of SDSS Stripe 82 Quasars as a Damped Random Walk}",
      journal = {\apj},
         year = 2010,
        month = oct,
       volume = {721},
       number = {2},
        pages = {1014-1033},
          doi = {10.1088/0004-637X/721/2/1014},
archivePrefix = {arXiv},
       eprint = {1004.0276},
 primaryClass = {astro-ph.CO},
       adsurl = {https://ui.adsabs.harvard.edu/abs/2010ApJ...721.1014M}
}

@article{vaughan2016false,
       author = {{Vaughan}, S. and {Uttley}, P. and {Markowitz}, A.~G. and {Huppenkothen}, D. and {Middleton}, M.~J. and {Alston}, W.~N. and {Scargle}, J.~D. and {Farr}, W.~M.},
        title = "{False periodicities in quasar time-domain surveys}",
      journal = {\mnras},
         year = 2016,
        month = sep,
       volume = {461},
       number = {3},
        pages = {3145-3152},
          doi = {10.1093/mnras/stw1412},
archivePrefix = {arXiv},
       eprint = {1606.02620},
 primaryClass = {astro-ph.IM},
       adsurl = {https://ui.adsabs.harvard.edu/abs/2016MNRAS.461.3145V}
}

@ARTICLE{kozowski2017limitationsdrw,
       author = {{Koz{\l}owski}, Szymon},
        title = "{Limitations on the recovery of the true AGN variability parameters using damped random walk modeling}",
      journal = {\aap},
         year = 2017,
        month = jan,
       volume = {597},
          eid = {A128},
        pages = {A128},
          doi = {10.1051/0004-6361/201629890},
archivePrefix = {arXiv},
       eprint = {1611.08248},
 primaryClass = {astro-ph.GA},
       adsurl = {https://ui.adsabs.harvard.edu/abs/2017A&A...597A.128K}
}

@ARTICLE{zu2013isquasardpw,
       author = {{Zu}, Ying and {Kochanek}, C.~S. and {Koz{\l}owski}, Szymon and {Udalski}, Andrzej},
        title = "{Is Quasar Optical Variability a Damped Random Walk?}",
      journal = {\apj},
         year = 2013,
        month = mar,
       volume = {765},
       number = {2},
          eid = {106},
        pages = {106},
          doi = {10.1088/0004-637X/765/2/106},
archivePrefix = {arXiv},
       eprint = {1202.3783},
 primaryClass = {astro-ph.CO},
       adsurl = {https://ui.adsabs.harvard.edu/abs/2013ApJ...765..106Z}
}

@article{graham2015systematic,
        author = {{Graham}, Matthew J. and {Djorgovski}, S.~G. and {Stern}, Daniel and {Drake}, Andrew J. and {Mahabal}, Ashish A. and {Donalek}, Ciro and {Glikman}, Eilat and {Larson}, Steve and {Christensen}, Eric},
        title = "{A systematic search for close supermassive black hole binaries in the Catalina Real-time Transient Survey}",
      journal = {\mnras},
         year = 2015,
        month = oct,
       volume = {453},
       number = {2},
        pages = {1562-1576},
          doi = {10.1093/mnras/stv1726},
archivePrefix = {arXiv},
       eprint = {1507.07603},
 primaryClass = {astro-ph.GA},
       adsurl = {https://ui.adsabs.harvard.edu/abs/2015MNRAS.453.1562G}
}

@article{graham2015possible,
       author = {{Graham}, Matthew J. and {Djorgovski}, S.~G. and {Stern}, Daniel and {Glikman}, Eilat and {Drake}, Andrew J. and {Mahabal}, Ashish A. and {Donalek}, Ciro and {Larson}, Steve and {Christensen}, Eric},
        title = "{A possible close supermassive black-hole binary in a quasar with optical periodicity}",
      journal = {\nat},
         year = 2015,
        month = feb,
       volume = {518},
       number = {7537},
        pages = {74-76},
          doi = {10.1038/nature14143},
archivePrefix = {arXiv},
       eprint = {1501.01375},
 primaryClass = {astro-ph.GA},
       adsurl = {https://ui.adsabs.harvard.edu/abs/2015Natur.518...74G}
}

@article{charisi2016population,
       author = {{Charisi}, M. and {Bartos}, I. and {Haiman}, Z. and {Price-Whelan}, A.~M. and {Graham}, M.~J. and {Bellm}, E.~C. and {Laher}, R.~R. and {M{\'a}rka}, S.},
        title = "{A population of short-period variable quasars from PTF as supermassive black hole binary candidates}",
      journal = {\mnras},
         year = 2016,
        month = dec,
       volume = {463},
       number = {2},
        pages = {2145-2171},
          doi = {10.1093/mnras/stw1838},
archivePrefix = {arXiv},
       eprint = {1604.01020},
 primaryClass = {astro-ph.GA},
       adsurl = {https://ui.adsabs.harvard.edu/abs/2016MNRAS.463.2145C}
}

@ARTICLE{liu2018asassnsmbhb,
       author = {{Liu}, Tingting and {Gezari}, Suvi and {Miller}, M. Coleman},
        title = "{Did ASAS-SN Kill the Supermassive Black Hole Binary Candidate PG1302-102?}",
      journal = {\apjl},
         year = 2018,
        month = may,
       volume = {859},
       number = {1},
          eid = {L12},
        pages = {L12},
          doi = {10.3847/2041-8213/aac2ed},
archivePrefix = {arXiv},
       eprint = {1803.05448},
 primaryClass = {astro-ph.HE},
       adsurl = {https://ui.adsabs.harvard.edu/abs/2018ApJ...859L..12L}
}

@ARTICLE{liu2019smbhbpanstarrs,
       author = {{Liu}, T. and {Gezari}, S. and {Ayers}, M. and {Burgett}, W. and {Chambers}, K. and {Hodapp}, K. and {Huber}, M.~E. and {Kudritzki}, R. -P. and {Metcalfe}, N. and {Tonry}, J. and {Wainscoat}, R. and {Waters}, C.},
        title = "{Supermassive Black Hole Binary Candidates from the Pan-STARRS1 Medium Deep Survey}",
      journal = {\apj},
         year = 2019,
        month = oct,
       volume = {884},
       number = {1},
          eid = {36},
        pages = {36},
          doi = {10.3847/1538-4357/ab40cb},
archivePrefix = {arXiv},
       eprint = {1906.08315},
 primaryClass = {astro-ph.HE},
       adsurl = {https://ui.adsabs.harvard.edu/abs/2019ApJ...884...36L}
}

@ARTICLE{witt2022periodicquasars,
       author = {{Witt}, Caitlin A. and {Charisi}, Maria and {Taylor}, Stephen R. and {Burke-Spolaor}, Sarah},
        title = "{Quasars with Periodic Variability: Capabilities and Limitations of Bayesian Searches for Supermassive Black Hole Binaries in Time-domain Surveys}",
      journal = {\apj},
         year = 2022,
        month = sep,
       volume = {936},
       number = {1},
          eid = {89},
        pages = {89},
          doi = {10.3847/1538-4357/ac8356},
archivePrefix = {arXiv},
       eprint = {2110.07465},
 primaryClass = {astro-ph.GA},
       adsurl = {https://ui.adsabs.harvard.edu/abs/2022ApJ...936...89W}
}

@ARTICLE{Chen+2024,
       author = {{Chen}, Yong-Jie and {Zhai}, Shuo and {Liu}, Jun-Rong and {Guo}, Wei-Jian and {Peng}, Yue-Chang and {Li}, Yan-Rong and {Songsheng}, Yu-Yang and {Du}, Pu and {Hu}, Chen and {Wang}, Jian-Min},
        title = "{Searching for quasar candidates with periodic variations from the Zwicky Transient Facility: results and implications}",
      journal = {\mnras},
         year = 2024,
        month = feb,
       volume = {527},
       number = {4},
        pages = {12154-12177},
          doi = {10.1093/mnras/stad3981},
archivePrefix = {arXiv},
       eprint = {2206.11497},
 primaryClass = {astro-ph.GA},
       adsurl = {https://ui.adsabs.harvard.edu/abs/2024MNRAS.52712154C}
}

@ARTICLE{zhu2020unambiguous,
       author = {{Zhu}, Xing-Jiang and {Thrane}, Eric},
        title = "{Toward the Unambiguous Identification of Supermassive Binary Black Holes through Bayesian Inference}",
      journal = {\apj},
         year = 2020,
        month = sep,
       volume = {900},
       number = {2},
          eid = {117},
        pages = {117},
          doi = {10.3847/1538-4357/abac5a},
archivePrefix = {arXiv},
       eprint = {2004.10944},
 primaryClass = {astro-ph.HE},
       adsurl = {https://ui.adsabs.harvard.edu/abs/2020ApJ...900..117Z}
}

@article{park2024self,
       author = {{Park}, Kevin and {Xin}, Chengcheng and {Davelaar}, Jordy and {Haiman}, Zoltan},
        title = "{Self-lensing flares from black hole binaries IV: the number of detectable shadows}",
      journal = {arXiv e-prints},
         year = 2024,
        month = sep,
          eid = {arXiv:2409.04583},
        pages = {arXiv:2409.04583},
          doi = {10.48550/arXiv.2409.04583},
archivePrefix = {arXiv},
       eprint = {2409.04583},
 primaryClass = {astro-ph.HE},
       adsurl = {https://ui.adsabs.harvard.edu/abs/2024arXiv240904583P}
}

@article{lomb1976least,
       author = {{Lomb}, N.~R.},
        title = "{Least-Squares Frequency Analysis of Unequally Spaced Data}",
      journal = {\apss},
         year = 1976,
        month = feb,
       volume = {39},
       number = {2},
        pages = {447-462},
          doi = {10.1007/BF00648343},
       adsurl = {https://ui.adsabs.harvard.edu/abs/1976Ap&SS..39..447L}
}

\begin{appendix}
%===============

%--------------------------------------
\section{Supporting figures and tables}
%--------------------------------------
\label{sec:supporting_material}

\subsubsection*{Section 2}

The ADQL query to retrieve the source identifiers of the initial AGN sample:

\begin{verbatim}
SELECT qso.source_id, qso.redshift_qsoc, 
vari.num_selected_g_fov, vari.time_duration_g_fov, 
spurious.spearman_corr_ipd_g_fov, 
spurious.spearman_corr_exf_g_fov
FROM gaiadr3.qso_candidates 
    AS qso
JOIN gaiadr3.vari_summary 
    AS vari USING (source_id)
JOIN gaiadr3.vari_spurious_signals 
    AS spurious USING (source_id)
WHERE qso.gaia_crf_source='t'
AND qso.redshift_qsoc > 0.1 
AND vari.num_selected_g_fov >= 40
AND vari.time_duration_g_fov >= 500
AND (spurious.spearman_corr_ipd_g_fov<0.8 OR 
spurious.scan_angle_model_ampl_sig_g_fov<6)
\end{verbatim}

\subsubsection*{Section 4}

\begin{figure}[t]
    \centering
    \includegraphics[width=0.48\textwidth]{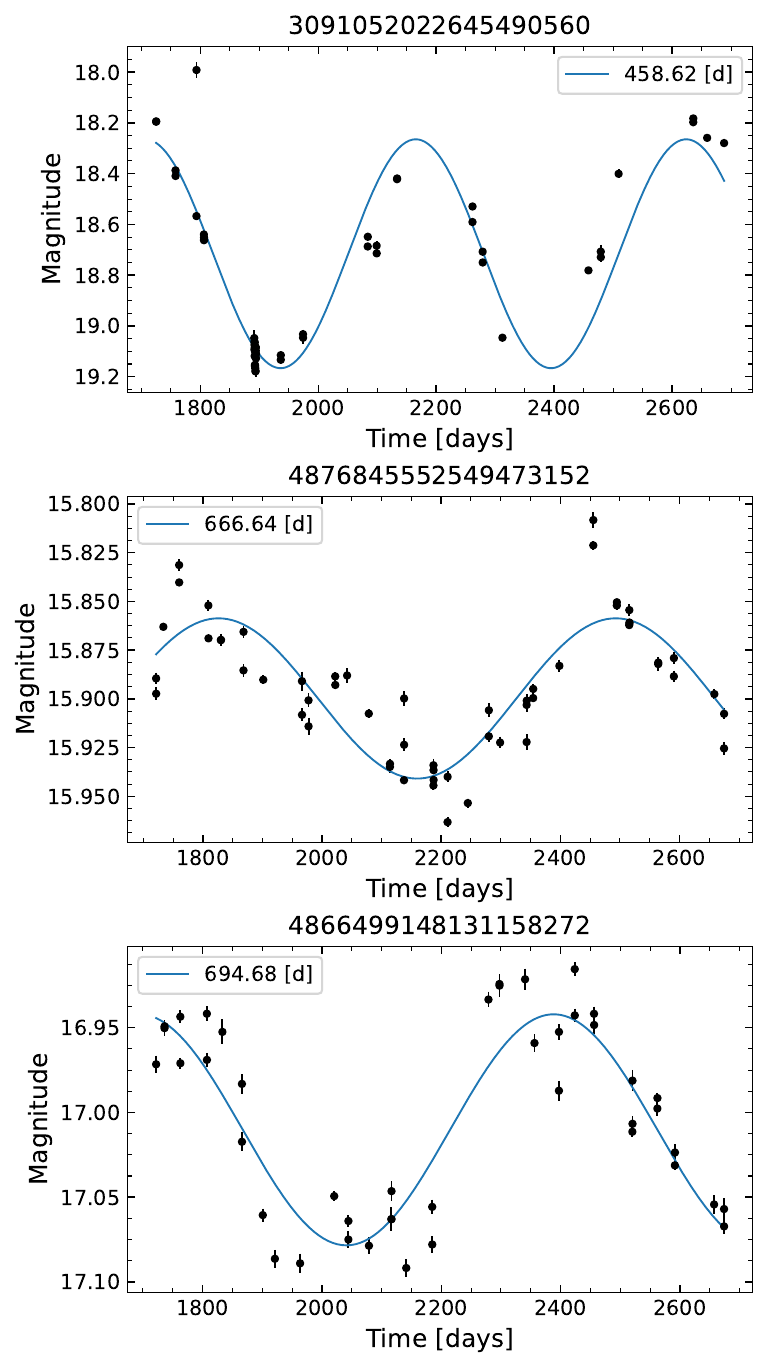}
    \caption{Light curves of three significant candidates with shortest dominant period. Only the top source exhibits a period shorter than $T/1.5$, covering approximately 2.1 cycles.}
    \label{fig:best_candidates}
\end{figure}

The 13 candidates that passed all filters are given in Table~\ref{tab:candidates} and the first three rows are shown in Figure~\ref{fig:best_candidates}.

\begin{table}[t]
\centering
\caption{Candidates that passed all filters sorted by their dominant periodicity.}
\label{tab:candidates}
\begin{tabular}{lccc}
\hline\hline
Source ID & RA [deg] & Dec [deg] & Period [d] \\
\hline
3091052022645490560 & $122.2934$ & $2.9302$ & $458.6$ \\
4876845552549473152 & $76.0793$ & $-29.7442$ & $666.6$ \\
4866499148131158272 & $69.2271$ & $-37.2608$ & $694.7$ \\
1474818072105266944 & $199.7536$ & $37.7372$ & $750.9$ \\
5623533760221472384 & $139.7787$ & $-35.6408$ & $895.9$ \\
5015127281684030336 & $21.6353$ & $-34.0981$ & $926.0$ \\
4394574219026344832 & $258.5679$ & $7.1703$ & $937.8$ \\
4947320945057622400 & $42.3668$ & $-43.8540$ & $948.9$ \\
6530809691273048960 & $359.5384$ & $-46.0836$ & $952.8$ \\
4935848026554944000 & $22.5051$ & $-42.6723$ & $968.8$ \\
4946491814507959424 & $36.8225$ & $-44.5571$ & $977.7$ \\
552258375370640640 & $79.9643$ & $76.9335$ & $998.9$ \\
3531556654642128896 & $170.7595$ & $-27.5012$ & $1013.0$ \\
\hline
\end{tabular}
\end{table}

\end{appendix}

\end{document}